\documentclass[reqno,11pt]{amsart}
\usepackage{amsmath, latexsym, amsfonts, amssymb, amsthm, amscd}
\usepackage{graphics,epsf,psfrag}
\numberwithin{equation}{section}

\newtheorem{theorem}{Theorem}[section]

\newtheorem{rem}[theorem]{Remark}

\newcommand{\ind}{\mathbf{1}}

\newcommand{\R}{\mathbb{R}}
\newcommand{\Z}{\mathbb{Z}}
\newcommand{\N}{\mathbb{N}}
\renewcommand{\tilde}{\widetilde}

\newcommand{\med}[1]{\left\langle #1\right\rangle}

\newcommand{\bP}{{\ensuremath{\mathbf P}} }
\newcommand{\bE}{{\ensuremath{\mathbf E}} }

\DeclareMathSymbol{\leqslant}{\mathalpha}{AMSa}{"36} 
\DeclareMathSymbol{\geqslant}{\mathalpha}{AMSa}{"3E} 
\DeclareMathSymbol{\eset}{\mathalpha}{AMSb}{"3F}     
\newcommand{\dd}{\,\text{\rm d}}             

\newcommand{\bbE}{{\ensuremath{\mathbb E}} }

\newcommand{\bbP}{{\ensuremath{\mathbb P}} }

\newcommand{\go}{\omega}

\makeatletter
\def\captionfont@{\footnotesize}
\def\captionheadfont@{\scshape}

\long\def\@makecaption#1#2{%
  \vspace{2mm} \setbox\@tempboxa\vbox{\color@setgroup
  \advance\hsize-6pc\noindent
  \captionfont@\captionheadfont@#1\@xp\@ifnotempty\@xp
  {\@cdr#2\@nil}{.\captionfont@\upshape\enspace#2}%
  \unskip\kern-6pc\par \global\setbox\@ne\lastbox\color@endgroup}%
  \ifhbox\@ne 
  \setbox\@ne\hbox{\unhbox\@ne\unskip\unskip\unpenalty\unkern}
  \ifdim\wd\@tempboxa=\z@ 
  \setbox\@ne\hbox to\columnwidth{\hss\kern-6pc\box\@ne\hss}
  tempboxa contained more than one line
  \setbox\@ne\vbox{\unvbox\@tempboxa\parskip\z@skip
  \noindent\unhbox\@ne\advance\hsize-6pc\par}%
\fi
  \ifnum\@tempcnta<64 
    \addvspace\abovecaptionskip
    \moveright 3pc\box\@ne
  \else 
    \moveright 3pc\box\@ne
    \nobreak
    \vskip\belowcaptionskip
  \fi
\relax
}
\makeatother
\def\writefig#1 #2 #3 {\rlap{\kern #1 truecm
\raise #2 truecm \hbox{#3}}}

\begin{document}

\title[Pinning models and replica coupling]{
A replica-coupling approach to disordered pinning models}

\author{Fabio Lucio Toninelli}
\address{\noindent Universit\'e de Lyon,\hfill\break
Laboratoire de Physique de l'Ecole Normale Sup\'erieure de Lyon, 
CNRS UMR 5672,\hfill\break 46 All\'ee d'Italie,
69364 Lyon, France
}
\email{fltonine@ens-lyon.fr}
\date{\today}

\begin{abstract}
We consider a renewal process $\tau=\{\tau_0,\tau_1,\ldots\}$ on the
integers, where the law of $\tau_i-\tau_{i-1}$ has a power-like tail
$\bP(\tau_i-\tau_{i-1}=n)=n^{-(\alpha+1)}L(n)$ with $\alpha\ge0$ and
$L(\cdot)$ slowly varying. We then assign a random, $n$-dependent
reward/penalty to the occurrence of the event that the site $n$
belongs to $ \tau$.  In such generality this class of problems
includes, among others, $(1+d)$-dimensional models of pinning of
directed polymers on a one-dimensional random defect,
$(1+1)$-dimensional models of wetting of disordered substrates, and
the Poland-Scheraga model of DNA denaturation. By varying the average
of the reward, the system undergoes a transition from a {\sl localized
phase} where $\tau$ occupies a finite fraction of $\N$ to a {\sl
delocalized phase} where the density of $\tau$ vanishes.  In absence
of disorder (i.e., if the reward is independent of $n$), the transition
is of first order for $\alpha>1$ and of higher order for
$\alpha<1$. Moreover, for $\alpha$ ranging from $1$ to $0$, the
transition ranges from first to infinite order. Presence of even an
arbitrarily small (but extensive) amount of disorder is known to
modify the order of transition as soon as $\alpha>1/2$
\cite{cf:GT_smooth}. In physical terms, disorder is relevant in this
situation, in agreement with the heuristic Harris criterion. On the
other hand, for $0<\alpha<1/2$ it has been proven recently by
K. Alexander \cite{cf:Ken} that, if disorder is sufficiently weak,
critical exponents are not modified by randomness: disorder is
irrelevant.  In this work, generalizing techniques which in the
framework of spin glasses are known as {\sl replica coupling} and {\sl
interpolation}, we give a new, simpler proof of the main results of
\cite{cf:Ken}.  Moreover, we (partially) justify a small-disorder
expansion worked out in \cite{cf:GB} for $\alpha<1/2$, showing that it
provides a free energy upper bound which improves the annealed one.
\end{abstract}

\maketitle

\section{Introduction}

Consider a (recurrent or transient) Markov chain $\{S_n\}_{n\ge0}$
started from a particular point, call it $0$ by convention, of the
state space $\Sigma$.  Assume that the distribution of the inter-arrival times
to the state $0$ has a power-like tail: if $\tau:=\{n\ge0:S_n=0\}$, we
require $\bP(\tau_i-\tau_{i-1}=n) \simeq n^{-\alpha-1}$ for $n$ large
(see Eq. \eqref{eq:K} below for precise definitions and
conditions). This is true, for instance, if $S$ is the simple
random walk (SRW) in $\Sigma=\Z^d$, in which case $\alpha=1/2$ for $d=1$ and
$\alpha=d/2-1$ for $d\ge2$. One may naturally think of
$\{(n,S_n)\}_{n\ge0}$ as a directed polymer configuration in
$\Sigma\times \N$. We want to model the situation where the polymer interacts
with the one-dimensional defect line $\{0\}\times \N$. To this purpose,
we introduce the Hamiltonian
\begin{eqnarray}
  \mathcal H_N(S)=-\sum_{n=1}^N \varepsilon_n \ind_{S_n=0}
\end{eqnarray}
which gives a reward (if $\varepsilon_n>0$) or a penalty
(if $\varepsilon_n<0$) to the occurrence of a polymer-line contact at
step $n$.  Typically, we have in mind the situation where
$\{\varepsilon_n\}_{n\in \N}$ is a sequence of IID (possibly
degenerate) random variables.  Let $h$ and $\beta^2 $ be the average
and variance of $\varepsilon_n$, respectively.  Varying $h$ at $\beta$
fixed, the system undergoes a phase transition: for $h>h_c(\beta)$ the
Boltzmann average of the contact fraction $\ell_N:=|\{1\le n\le
N:S_n=0\}|/N$ converges almost surely to a positive constant, call it
$\ell(\beta,h)$, for $N\to\infty$ (localized phase), while for
$h<h_c(\beta)$ it converges to zero (delocalized phase).  Models of
this kind are employed in the physics literature to describe, for
instance, the interaction of $(1+1)$-dimensional interfaces with
disordered walls \cite{cf:Derrida}, of flux lines with columnar
defects in type-II superconductors \cite{cf:NV}, and the DNA
denaturation transition in the Poland-Scheraga approximation
\cite{cf:Cule}.

 In absence of disorder ($\beta=0$) it is known that the transition is
 of first order ($\ell(0,h)$ has a discontinuity at $h_c(0)$) if
 $\alpha>1$, while for $0\le\alpha<1$ the transition is continuous: in
 particular, $\ell(0,h)$ vanishes like
 $(h-h_c(0))^{(1-\alpha)/\alpha}$ for $h\searrow h_c(0)$ if
 $0<\alpha<1$ and faster than any power of $(h-h_c(0))$ if
 $\alpha=0$. For $\alpha=1$, finally, transition can be either
 continuous or discontinuous, depending on the slowly varying function
 $L(\cdot)$ in \eqref{eq:K}.  An interesting question concerns the effect
 of disorder on the nature of the transition. In terms of the
 non-rigorous Harris criterion, disorder is believed to be irrelevant
 for $\alpha<1/2$ and relevant for $\alpha>1/2$, where ``relevance''
 refers to the property of changing the critical exponents. The
 question of disorder relevance in the (so called ``marginal'') case
 $\alpha=1/2$ is not settled yet, even on heuristic grounds. Recently,
 rigorous methods have allowed to put this belief on firmer ground. In
 particular, in Refs.  \cite{cf:GT_smooth}-\cite{cf:GT_prl} it was
 proved that, for every $\beta>0$, $\alpha\ge0$ and $h$ 
 sufficiently close to (but larger than) $h_c(\beta)$, one
 has $\ell(\beta,h)\le (1+\alpha) c(\beta)(h-h_c(\beta))$, for some
 $c(\beta)<\infty$.  This result, compared with the critical behavior
 mentioned above of the non-random model, proves relevance of disorder
 for $\alpha>1/2$, since $1>(1-\alpha)/\alpha$. On the other hand, in
 a recent remarkable work K. Alexander showed \cite{cf:Ken} that the
 opposite is true for $0<\alpha<1/2$: if disorder is sufficiently
 weak, $\ell(\beta,h)$ vanishes like $
 (h-h_c(\beta))^{(1-\alpha)/\alpha}$ as in the homogeneous model.
 Moreover, the critical point $h_c(\beta)$ coincides, always for
 $\beta$ small and $0<\alpha<1/2$, with the critical point
 $h^a_c(\beta)$ of the corresponding {\sl annealed model} (cf. Section
\ref{sec:mod}). Always in
\cite{cf:Ken}, for $1/2\le \alpha<1$  it was
proven that the ratio $h_c(\beta)/h_c^a(\beta)$ converges to $1$ for
$\beta\searrow0$. This, on the other hand, is expected to be false for
$\alpha>1$.

The purpose of this work is twofold. Firstly, we present a method
which allows to re-obtain the main results of \cite{cf:Ken} in a
simpler way. 
Secondly, we show that the well known inequality between quenched and
annealed free energies is strict as soon as the annealed model is
localized and $\beta>0$. Moreover, we prove that a small-disorder
expansion for the quenched free energy, worked out in \cite{cf:GB} for
$0\le\alpha<1/2$, provides at least a free energy upper bound.

As far as the first point is concerned, our strategy is a generalization
of techniques which in the domain of mean field spin glasses are known
as {\sl replica coupling}
\cite{cf:T_ap} \cite{cf:GuT_quadr} and {\sl interpolation}. These methods
had a remarkable impact on the understanding of spin glasses in recent
years (see, e.g., \cite{cf:GuT_cmp},
\cite{cf:Gu}, \cite{cf:Ass},
\cite{cf:T}). In particular the ``quadratic replica coupling'' method,
introduced in \cite{cf:GuT_quadr}, gives a very efficient control of
the Sherrington-Kirkpatrick model at high temperature ($\beta$ small),
i.e., for weak disorder, which is the same situation we are after
here. Our method is not unrelated to that of
\cite{cf:Ken}: the two share the idea that the basic object to look
at is the law of the intersection of two independent copies of the
renewal $\tau$. However, our strategy
allows  to bypass the need of performing refined second-moment
computations on a suitably truncated partition function 
as in \cite{cf:Ken} and gives, in
the case of Gaussian disorder, particularly neat proofs.

In the rest of the paper, we will forget the polymer-like interpretation
and the Markov chain structure, and define the model directly starting from
the process $\tau$ of the ``returns to $0$''.

\section{Model and results}
\label{sec:mod}

Consider a {\sl recurrent} renewal sequence
$0=\tau_0<\tau_1<\ldots$ where $\{\tau_i-\tau_{i-1}\}_{i\ge0}$ are
integer-valued IID random variables with law
\begin{eqnarray}
\label{eq:K}
K(n):= \bP(\tau_1=n)= \frac{L(n)}{n^{1+\alpha}}\;\;\;\;\;\forall n\in\N.
\end{eqnarray}
We assume that the function $L(\cdot)$ is slowly varying at infinity
\cite{cf:bingh}, $\alpha\ge 0$ and $\sum_{n\in\N}K(n)=1$. Recall that a
slowly varying function $L(\cdot)$ is a positive function $(0,\infty)\ni
x\to L(x)\in (0,\infty)$ such that, for every $r>0$,
\begin{eqnarray}
\label{eq:slow}
\lim_{x\to\infty}\frac{L(rx)}{L(x)}=1.
\end{eqnarray}
We denote by $\bE$ the expectation
on $\tau:=\{\tau_i\}_{i\ge0}$  and we put for notational simplicity
$\delta_n:= \ind_{n\in\tau}$, where $\ind_A$ is the indicator of the 
event $A$.

We define, for $\beta\ge0$ and $h\in \mathbb R$, the quenched free
energy as
\begin{eqnarray}
\label{eq:F}
  F(\beta,h)= \lim_{N\to\infty}F_N(\beta,h):=
\lim_{N\to\infty}\frac1N\bbE \log \bE \left(
e^{\sum_{n=1}^N(\beta\go_n+h)\delta_n}
\,\delta_N
\right)
\end{eqnarray}
where $\{\go_n\}_{n\in \N}$ are IID centered random variables with 
finite second moment, law
denoted by $\bbP$ and corresponding expectation $\bbE$, and normalized
so that $\bbE \,\go_1^2=1$. In this work, we restrict to the case
where disorder has a Gaussian distribution: $\go_1\stackrel d= \mathcal
N(0,1)$. Some degree of generalization is possible: for instance,
results and proofs can be extended to the situation where $\go_n$ are
IID bounded random variables.

The existence of the $N\to\infty$ limit in \eqref{eq:F} is well known,
see for instance
\cite[Section 4.2]{cf:GB}. The limit actually exists, and is almost-surely
equal to $F(\beta,h)$, even omitting the disorder average $\bbE$ in
\eqref{eq:F}.
We point out that, by superadditivity, for every $N\in\N$
\begin{eqnarray}
\label{eq:superadd}
F_N(\beta,h)\le F(\beta,h)
\end{eqnarray}
and that, from Jensen's inequality,
\begin{eqnarray}
\label{eq:jensen}
F_N(\beta,h)\le F_N^a(\beta,h):=\frac1N\log \bbE\, \bE\left(
e^{\sum_{n=1}^N(\beta\go_n+h)\delta_n}
\delta_N
\right)= F_N(0,h+\beta^2/2).
\end{eqnarray}
(If disorder is not Gaussian, $\beta^2/2$ is replaced by
$\log \bbE \exp(\beta \go_1)$.)
$F^a_N(\beta,h)$ is known as the (finite-volume) {\sl annealed} free
energy which, as  \eqref{eq:jensen} shows, coincides with the free
energy of the homogeneous model ($\beta=0$) for a shifted value of $h$.  The
limit free energy $F(\beta,h)$ would not change (see, e.g., \cite[Remark
1.1]{cf:GT_smooth}) if the factor $\delta_N$ were omitted in
\eqref{eq:F}, i.e., if the boundary condition $\{N\in\tau\}$ were replaced
by a free boundary condition at $N$. However, in that case exact
superadditivity, and \eqref{eq:superadd}, would not hold.

Another well-established fact is that $F(\beta,h)\ge0$ (cf. for
instance
\cite{cf:GT_smooth}), which allows the  definition of
the critical point, for a given $\beta\ge0$, as
$h_c(\beta):=\sup\{h\in \R:F(\beta,h)=0\}$.
Note that Eq. \eqref{eq:jensen} implies  $h_c(\beta)\ge h^a_c(\beta)
:=\sup\{h\in\R:F^a(\beta,h)=0\}=h_c(0)-\beta^2/2$. For obvious reasons,
$h_c^a(\beta)$ is referred to as the {\sl annealed critical point}.
Concerning upper bounds for $h_c(\beta)$, already before \cite{cf:Ken}
it was known (see \cite{cf:AS} and \cite[Theorem 5.2]{cf:GB}) that
$h_c(\beta)<h_c(0)$ for every $\beta>0$.
To make a link with the discussion in the introduction, note that
the contact fraction $\ell(\beta,h)$ is just $\partial_h F(\beta,h)$.

With the exception of Theorem \ref{th:rs}, we will consider from now
on only the values $0<\alpha<1$, as in \cite{cf:Ken}, in which case
$\tau$ is null-recurrent under $\bP$. For the homogeneous system it is
known \cite[Theorem 2.1]{cf:GB} that $F(0,h)=0$ if $h\le 0$, while for
$h>0$
\begin{eqnarray}
\label{eq:asint}
  F(0,h)=h^{1/\alpha}\tilde L(1/h).
\end{eqnarray}
$\tilde L(\cdot)$ is a slowly varying function satisfying
\begin{eqnarray}
\tilde
L(1/h)=\left(\frac{\alpha}{\Gamma(1-\alpha)}\right)^{1/{\alpha}}
h^{-1/\alpha}R_\alpha(h),
\end{eqnarray}
and $R_\alpha(\cdot)$ is asymptotically equivalent to the inverse of the
map $x\mapsto x^{\alpha}L(1/x)$. The fact that $\tilde L(\cdot)$ is slowly
varying follows from \cite[Theorem 1.5.12]{cf:bingh}.  In particular,
notice that $h_c(0)=0$, so that $h^{a}_c(\beta)=-\beta^2/2$.

We want to prove first of all that, if $0<\alpha<1/2$ and
$\beta$ is sufficiently small (i.e., if
 disorder is sufficiently weak), $h_c(\beta)=h^a_c(\beta)$.
Keeping in mind that
  $F^a(\beta,h_c^a(\beta)+\Delta)=F(0,\Delta)$, this is an immediate
 consequence of
\begin{theorem}
\label{th:alpha<32}
  Assume that either $0<\alpha<1/2$ or that $\alpha=1/2$ and
$\sum_{n\in \N}n^{-1}L(n)^{-2}<\infty$. Then, for every
$\epsilon>0$ there exist $\beta_0(\epsilon)>0$ and
$\Delta_0(\epsilon)>0$ such that, for every $\beta\le
\beta_0(\epsilon)$ and $0<\Delta<\Delta_0(\epsilon)$, one has
\begin{eqnarray}
\label{eq:risulta1}
(1-\epsilon)F(0,\Delta)\, \le \, F(\beta,h_c^a(\beta)+\Delta)\le
F(0,\Delta).
\end{eqnarray}
\end{theorem}
\medskip

In view of \cite[Theorem 2.1]{cf:GT_smooth}, the same cannot hold for
$\alpha>1/2$. However, one has:
 \begin{theorem}
\label{th:a322}
   Assume that $1/2<\alpha<1$. There exists a slowly
varying function $ \check L(\cdot)$ and, for every $\epsilon>0$, constants
$a_1(\epsilon)< \infty$ and $\Delta_0(\epsilon)>0$ such that, if
\begin{eqnarray}
\label{eq:condiz}
a_1(\epsilon)\beta^{2\alpha/(2\alpha-1)}\check L (1/\beta)
\, \le\,  \Delta \,
\,
\le\,  \Delta_0(\epsilon),
\end{eqnarray}
the inequalities \eqref{eq:risulta1} hold.  
\end{theorem}
\medskip

As already pointed out in \cite{cf:Ken}, since $2\alpha/(2\alpha-1)>2$
Theorem \ref{th:a322} shows in particular that
\begin{eqnarray}
\lim_{\beta\searrow 0}\frac{h_c(\beta)}{h^a_c(\beta)}=1.
\end{eqnarray}
On the other hand, it is unknown whether there exist non-zero values
of $\beta$ such that the equality $h_c(\beta)=h_c^a(\beta)$ holds,
for $1/2<\alpha<1$.
\begin{rem}\rm
The lower bound we obtain for $F(\beta,h)$ (and, as a consequence, for
$h^a_c(\beta)-h_c(\beta)$) in Theorem \ref{th:a322} differs from the
analogous one of \cite[Theorem 3]{cf:Ken} only in the form of the
slowly varying function $\check L(\cdot)$ (an explicit choice of $\check
L(\cdot)$ can be extracted from Eq. \eqref{eq:horr} below). In general,
our $\check L(\cdot)$ is larger due to the logarithmic
denominator in \eqref{eq:horr}. However, the important point is that
the exponent of $\beta$ in \eqref{eq:condiz} agrees with that in the
analogous condition of \cite[Theorem 3]{cf:Ken}.  Indeed, with the
conventions of \cite{cf:Ken} (which amount to replacing everywhere $h$
by $\beta h$ and $\alpha$ by $c-1$), the exponent
$2\alpha/(2\alpha-1)= 1+1/(2\alpha-1)$ would be instead
$1/(2\alpha-1)=1/(2c-3)$, as in \cite[Theorem 3]{cf:Ken}.
\end{rem}

Finally, for the ``marginal case''  we have
\begin{theorem}
\label{th:a32}
   Assume that $\alpha=1/2$ and $\sum_{n\in \N}n^{-1}L(n)^{-2}=\infty$.
Let $\ell(\cdot)$ be the slowly varying function (diverging at infinity)
defined by
\begin{eqnarray}
\sum_{n=1}^N \frac 1{n L(n)^2}\stackrel{N\to\infty}{\sim}\ell(N).
\end{eqnarray}
For every $\epsilon>0$ there exist constants
$a_2(\epsilon)<\infty$ and $\Delta_0(\epsilon)>0$ 
such that, if $0<\Delta\le \Delta_0(\epsilon)$
and if the condition
\begin{eqnarray}
\label{eq:condiz3}
\frac1{\beta^2}\ge a_2(\epsilon)\,\ell\left(\frac{a_2(\epsilon)|\log 
F(0,\Delta)|}
{F(0,\Delta)
}\right)
\end{eqnarray}
is verified, then Eq. \eqref{eq:risulta1} holds.
\end{theorem}
\begin{rem}\rm
Note that, thanks to Theorem \ref{th:a32} and the property of slow
variation of $\ell(\cdot)$, the difference $
h_c(\beta)-h^a_c(\beta)$ vanishes faster than any power of $
\beta$, for $\beta\searrow0$. Again, it is unknown whether $h_c(\beta)=
h_c^a(\beta)$ for some $\beta>0$.

In general, our condition \eqref{eq:condiz3} is different from the one
in the analogous Theorem 4 of \cite{cf:Ken}, due to the presence of
the factor $|\log F(0,\Delta)|$ in the argument of $\ell(\cdot)$.
However, for many ``reasonable'' and physically interesting choices of
$L(\cdot)$ in
\eqref{eq:K}, the two results are equivalent. In particular, if $\bP$
is the law of the returns to zero of the SRW $\{S_n\}_{n\ge0}$ in one
dimension, i.e. $\tau=\{n\ge0:S_{2n}=0\}$, in which case $L(\cdot)$ and
$\tilde L(\cdot)$ are asymptotically constant and $\ell(N)\sim a_3 \log
N$, one sees easily that \eqref{eq:condiz3} is verified as soon as
\begin{eqnarray}
\Delta\ge a_4(\epsilon)e^{-\frac{a_5(\epsilon)}{\beta^2}},
\end{eqnarray}
which is the same condition which was found in \cite{cf:Ken}.
Another case where Theorem \ref{th:a32} and \cite[Theorem
4]{cf:Ken} are equivalent is when $L(n)\sim a_6(\log n)^{(1-\gamma)/2}$ for
$\gamma>0$, in which case $\ell(N)\sim a_{7}(\log N)^{\gamma}$.
\end{rem}

\medskip

While we  focused up to now on free energy lower bounds, one may
wonder whether it is possible to improve the Jensen upper bound 
\eqref{eq:jensen}. For $\alpha>1/2$ it follows from 
\cite{cf:GT_smooth} 
that $F(\beta,h)<F^a(\beta,h)$ as soon as $\beta$ is positive and
$h-h^a_c(\beta)$ is positive and small. We conclude this section with
a theorem which generalizes this result to arbitrary $\alpha$ and
$h>h^a_c(\beta)$, and which justifies (as an upper bound) a
small-$\beta$ expansion worked out in \cite[Section 5.5]{cf:GB}.

\begin{theorem}
\label{th:rs}
For every $\beta>0$, $\alpha\ge0$ and $\Delta>0$
\begin{eqnarray}
\label{eq:RS}
F(\beta,h^a_c(\beta)+\Delta)\le \inf_{0\le q\le \Delta/\beta^2}
\left(\frac{\beta^2q^2}2+F(0,\Delta-\beta^2q)\right)<F(0,\Delta).
\end{eqnarray}
As a consequence, for $0\le\alpha<1/2$ there exist $\beta_0>0$ and
$\Delta_0>0$ such that
\begin{eqnarray}
\label{eq:svil}
F(\beta,h^a_c(\beta)+\Delta)\le F(0,\Delta)-\frac{\beta^2}2
\left(\partial_\Delta F(0,\Delta)\right)^2(1+O(\beta^2))
\end{eqnarray}
for $\beta\le \beta_0$ and $\Delta\le \Delta_0$, where $O(\beta^2)$ is
independent of $\Delta$.
\end{theorem}
 The first
inequality in \eqref{eq:RS} is somewhat reminiscent of the
``replica-symmetric'' free energy upper bound \cite{cf:Gu_rs} for the
Sherrington-Kirkpatrick model.

The reader who wonders why we stopped at order $\beta^2$ in the
``expansion'' \eqref{eq:svil} should look at Remark \ref{rem:why?} 
below.  Note that, in view of Eqs. \eqref{eq:asint} and
\eqref{eq:L()}, $(\partial_\Delta F(0,\Delta))^2\ll F(0,\Delta)$ if
$\Delta$ is small and $\alpha<1/2$.  Observe also that, for $\alpha>
1/2$ and $\beta,\Delta$ small, taking the infimum in \eqref{eq:RS}
gives nothing substantially better than just choosing
$q=\Delta/\beta^2$, from which $F(\beta,h^a_c(\beta)+\Delta)\le
\Delta^2/(2\beta^2)$; essentially the same bound (with an extra factor
$(1+\alpha)$ in the right-hand side) is however already implied by
\cite[Theorem 2.1]{cf:GT_smooth} (see also
\cite[Remark 5.7]{cf:GB}).
\begin{rem}\rm
  As a general remark, we  emphasize that the assumption of
  recurrence for $\tau$, i.e.,  $\sum_{n\in\N}K(n)=1$ is
  by no means a restriction. Indeed, as has been observed several
  times in the literature (including \cite{cf:GT_smooth} and
\cite{cf:Ken}), if $\Sigma_K:=\sum_{n\in\N}K(n)<1$ one can
  define $\tilde K(n):=K(n)/\Sigma_K$, and of course the renewal
  $\tau$ with law $\tilde\bP(\tau_1=n)=\tilde K(n)$ is recurrent. Then,
it is immediate to realize from definition \eqref{eq:F} that
\begin{eqnarray}
F(\beta,h)=\tilde F(\beta,h+\log \Sigma_K),
\end{eqnarray}
$\tilde F$ being the free energy of the model defined as in
\eqref{eq:F} but with $\bP$ replaced by $\tilde\bP$.
In particular, $h^a_c(\beta)=-\log \Sigma_K-\beta^2/2$.
 Theorems \ref{th:alpha<32}-\ref{th:rs} are therefore transferred
 with obvious changes to this situation.

This observation allows to apply the results, for instance, to the
case where we consider the SRW $\{S_n\}_{n\ge0}$ in $\Z^3$, and we let
$\bP$ be the law of $\tau:=\{n\ge0: S_{2n}=0\}$, i.e., the law of its
returns to the origin.  In this case, assumption \eqref{eq:K} holds
with $\alpha=1/2$, $L(\cdot)$ asymptotically constant and, due to
transience, $\Sigma_K<1$. The same is true if $\{S_n\}_{n\ge0}$ is the
SRW on $\Z$, conditioned to be non-negative.
\end{rem}

\section{Proofs}

{\sl Proof of Theorem \ref{th:alpha<32}}.  In view of
Eq. \eqref{eq:jensen}, we have to prove only the first inequality in
\eqref{eq:risulta1}.  This is based on an adaptation of the {\sl
quadratic replica coupling method} of \cite{cf:GuT_quadr}, plus ideas suggested
by \cite{cf:Ken}. Let $\Delta>0$ and start from the identity
\begin{eqnarray}
\label{eq:identity}
  F(\beta,-\beta^2/2+\Delta)
=F(0,\Delta)+\lim_{N\to\infty}R_{N,\Delta}(\beta)
\end{eqnarray}
where
$$
R_{N,\Delta}(\beta):=
\frac1N\bbE
\log \med{e^{\sum_{n=1}^N(\beta\go_n-\beta^2/2)\delta_n}}_{N,\Delta},
$$
and, for a $\bP$-measurable function $f(\tau)$,
\begin{eqnarray}
  \langle f\rangle_{N,\Delta}:=\frac{\bE\left(e^{\Delta\sum_{n=1}^N\delta_n}
f\,
\delta_N
\right)}{\bE\left(e^{\Delta\sum_{n=1}^N\delta_n}
\delta_N
\right)}.
\end{eqnarray}
Via the Gaussian integration by parts formula
\begin{eqnarray}
\label{eq:byparts}
  \bbE \left(\go f(\go)\right)=\bbE f'(\go),
\end{eqnarray}
valid (if $\go$ is a standard Gaussian random variable $\mathcal N(0,1)$) for
every differentiable function $f(\cdot)$ such that
$\lim_{|x|\to\infty}\exp(-x^2/2)f(x)=0$ , one finds for $0<t<1$:
\begin{eqnarray}
\label{eq:ipp1}
  \frac{\dd}{\dd t}R_{N,\Delta}(\sqrt t\beta)=-\frac{\beta^2}{2N}
\sum_{m=1}^N\bbE\left\{ \left(\frac{\med
{\delta_m\,e^{\sum_{n=1}^N(\beta\sqrt t\go_n-t\beta^2/2)\delta_n}}_{N,\Delta}}
{\med
{e^{\sum_{n=1}^N(\beta\sqrt t\go_n-t\beta^2/2)\delta_n}}_{N,\Delta}}\right)^2
\right\}.
\end{eqnarray}
Define also, for $\lambda\ge0$,
\begin{eqnarray}
 \psi_{N,\Delta}(t,\lambda,\beta)&:=&
\frac1{2N}\bbE \log
\med{e^{H_N(t,\lambda,\beta;\tau^{(1)},\tau^{(2)})}}_{N,\Delta}^{\otimes 2}
\\\nonumber
&:=&\frac1{2N}\bbE \log
\med{
e^{\sum_{n=1}^N(\beta\sqrt t \go_n-t\beta^2/2)(\delta^{(1)}_n+\delta^{(2)}_n)
+\lambda\beta^2\sum_{n=1}^N
\delta^{(1)}_n\delta^{(2)}_n}}_{N,\Delta}^{\otimes 2}
\end{eqnarray}
where the product measure $\med{\cdot}_{N,\Delta}^{\otimes 2}$ acts on
the pair $(\tau^{(1)},\tau^{(2)})$, while
$\delta^{(i)}_n:=\ind_{n\in\tau^{(i)}}$.  Note that
$\psi_{N,\Delta}(t,\lambda,\beta)$ actually depends on $(t,\lambda,
\beta)$ only through the two
combinations $\beta^2t$ and $\beta^2\lambda$.  We add also that the
introduction of the parameter $t$, which could in principle be
avoided, allows for more natural expressions in the formulas which
follow. One has immediately
\begin{eqnarray}
   \psi_{N,\Delta}(0,\lambda,\beta)=
\frac1{2N}\log \med{
e^{\lambda\beta^2\sum_{n=1}^N
\delta^{(1)}_n\delta^{(2)}_n}}_{N,\Delta}^{\otimes 2}
\end{eqnarray}
and
\begin{eqnarray}
\label{eq:310}
 \psi_{N,\Delta}(t,0,\beta)=R_{N,\Delta}(\sqrt t\beta).
\end{eqnarray}
Again via integration by parts,
\begin{eqnarray}
\frac{\dd}{\dd t}\psi_{N,\Delta}(t,\lambda,\beta)&=&\frac{\beta^2}{2N}
\sum_{m=1}^N\bbE \frac{\med{\delta^{(1)}_m\delta^{(2)}_m
e^{H_N(t,\lambda,\beta;\tau^{(1)},\tau^{(2)})}}_{N,\Delta}^{\otimes
2} }{\med{e^{H_N(t,\lambda,\beta;\tau^{(1)},\tau^{(2)})}
}_{N,\Delta}^{\otimes 2}}\\\nonumber &&-
\frac{\beta^2}{4N}\sum_{m=1}^N\bbE
\left\{\left(\frac{\med{(\delta^{(1)}_m+\delta^{(2)}_m)
e^{H_N(t,\lambda,\beta;\tau^{(1)},\tau^{(2)})}}_{N,\Delta}^{\otimes
2} }{\med{e^{H_N(t,\lambda,\beta;\tau^{(1)},\tau^{(2)})}
}_{N,\Delta}^{\otimes 2}} \right)^2\right\}\\\nonumber &&\le
\frac{\beta^2}{2N}\bbE
\sum_{m=1}^N\frac{\med{\delta^{(1)}_m\delta^{(2)}_m
e^{H_N(t,\lambda,\beta;\tau^{(1)},\tau^{(2)})}}_{N,\Delta}^{\otimes
2} }{\med{e^{H_N(t,\lambda,\beta;\tau^{(1)},\tau^{(2)})}
}_{N,\Delta}^{\otimes 2}}= \frac{\dd}{\dd
\lambda}\psi_{N,\Delta}(t,\lambda,\beta),
\end{eqnarray}
so that, for every $0\le t\le 1$ and $\lambda$,
\begin{eqnarray}
\label{eq:daflusso}
\psi_{N,\Delta}(t,\lambda,\beta)\le \psi_{N,\Delta}(0,\lambda+t,\beta).
\end{eqnarray}
Going back to Eq. \eqref{eq:ipp1}, using convexity and monotonicity
of $\psi_{N,\Delta}
(t,\lambda,\beta)$ with respect to $\lambda$ and \eqref{eq:310}, one finds
\begin{eqnarray}
&&   \frac{\dd}{\dd
t}\left(-R_{N,\Delta}(\sqrt t\beta)\right)=\frac{\dd}{\dd\lambda}
\left.\psi_{N,\Delta}(t,\lambda,\beta)\right|_{\lambda=0}
\le\frac{\psi_{N,\Delta}(t,2-t,\beta)-R_{N,\Delta}(\sqrt t\beta)}{2-t}
\\\nonumber
&&\le \psi_{N,\Delta}(0,2,\beta)-R_{N,\Delta}(\sqrt t\beta),
\end{eqnarray}
where in the last inequality we used \eqref{eq:daflusso} and the fact
that $2-t\ge1$.  Integrating this differential inequality between $0$
and $1$ and observing that $R_{N,\Delta}(0)=0,$ one has
\begin{eqnarray}
\label{eq:integrating}
  0\le -R_{N,\Delta}(\beta)\le (e-1)\psi_{N,\Delta}(0,2,\beta).
\end{eqnarray}
Now we estimate
\begin{equation}
\label{eq:315}
\begin{split}
\psi_{N,\Delta}(0,2,\beta)\, &=\,
-F_N(0,\Delta)+
\frac1{2N}\log \bE^{\otimes 2}\left
(e^{2\beta^2\sum_{n=1}^N\delta^{(1)}_n\delta^{(2)}_n+\Delta
\sum_{n=1}^N(\delta^{(1)}_n+\delta^{(2)}_n)}\delta^{(1)}_N\delta^{(2)}_N
\right)
\\
&\le\,  -F_N(0,\Delta)+\frac{F_N(0,q\Delta)}q+
\frac1{2Np}\log \bE^{\otimes 2} 
\left(e^{2p\beta^2\sum_{n=1}^N\delta^{(1)}_n\delta^{(2)}_n}
\right)
\end{split}
\end{equation}
where we used H\"older's inequality and $p,q$ (satisfying $1/p+1/q=1$)
are to be determined. One finds then
\begin{multline}
\limsup_{N\to\infty}
  \psi_{N,\Delta}(0,2,\beta)\le \limsup_{N\to\infty}
\frac1{2Np}\log
\bE^{\otimes 2} \left(e^{2p\beta^2\sum_{n=1}^{N}\delta^{(1)}_n\delta^{(2)}_n}
\right)\\
+F(0,\Delta)\left(\frac1q \frac{F(0,q\Delta)}{F(0,\Delta)}-1\right).
\end{multline}
But we know from \eqref{eq:asint} and the property
\eqref{eq:slow} of slow variation that, for every $q>0$,
\begin{eqnarray}
  \lim_{\Delta\searrow 0}\frac{F(0,q\Delta)}{F(0,\Delta)}=q^{1/\alpha}.
\end{eqnarray}
Therefore, choosing $q=q(\epsilon)$ sufficiently close to (but not
equal to) $1$ and $\Delta_0(\epsilon)>0$ sufficiently small one has,
uniformly on $\beta\ge0$ and on $0<\Delta\le \Delta_0(\epsilon)$,
\begin{eqnarray}
\label{eq:100}
 \limsup_{N\to\infty}\psi_{N,\Delta}(0,2,\beta)\le \frac{\epsilon}{e-1}
 F(0,\Delta)+
\limsup_{N\to\infty}\frac1{2Np(\epsilon)}\log
\bE^{\otimes 2} \left(e^{2p(\epsilon)\beta^2\sum_{n=1}^{N}\delta^{(1)}_n\delta^{(2)}_n}
\right).
\end{eqnarray}
Of course, $p(\epsilon)=q(\epsilon)/(q(\epsilon)-1)<\infty$ as long
as $\epsilon>0$.  Finally, we observe that under the assumptions of
the theorem, the renewal $\tau^{(1)}\cap \tau^{(2)}$ is transient
under the law $\bP^{\otimes 2}$.
Indeed, if $0<\alpha<1/2$ or if $\alpha=1/2$ and
$\sum_{n\in \N}n^{-1}L(n)^{-2}<\infty$ one has
\begin{eqnarray}
\label{eq:ricorrenza}
  \bE^{\otimes 2}\left(\sum_{n\ge1}\ind_{n\in\tau^{(1)}\cap\tau^{(2)}}\right)=\sum_{n\ge1}
\bP(n\in\tau)^2<\infty
\end{eqnarray}
since, as proven in \cite{cf:doney},
\begin{eqnarray}
\label{eq:doney}
\bP(n\in\tau)\stackrel{n\to\infty}\sim\frac{C_\alpha}{L(n)n^{1-\alpha}}:=
\frac{\alpha\sin(\pi \alpha)}\pi
\frac1{L(n)n^{1-\alpha}}.
\end{eqnarray}
Actually, Eq. \eqref{eq:doney} holds more generally for $0<\alpha<1$.

Therefore, there exists $\beta_1>0$ such that
\begin{eqnarray}
\sup_N \bE^{\otimes 2} \left(e^{2p(\epsilon)\beta^2\sum_{n=1}^{N}
\delta^{(1)}_n\delta^{(2)}_n}\right)<\infty
\end{eqnarray}
for every $\beta^2p(\epsilon)\le \beta_1^2$.  Together with
\eqref{eq:100} and \eqref{eq:identity}, this implies
\begin{eqnarray}
F(\beta,-\beta^2/2+\Delta)\ge (1-\epsilon)F(0,\Delta)
\end{eqnarray}
as soon as $\beta^2\le \beta_0^2(\epsilon):=\beta^2_1/p(\epsilon)$.
\hfill$\Box$

\medskip

{\sl Proof of Theorem \ref{th:a322}.} In what follows we assume that
$\Delta$ is sufficiently small so that $F(0,\Delta)<1$.  Let
$N=N(\Delta):=c|\log F(0,\Delta)|/F(0,\Delta)$ with $c>0$.  By
Eq. \eqref{eq:superadd} we have, in analogy with
\eqref{eq:identity},
 \begin{eqnarray} \label{eq:identity2} F(\beta,-\beta^2/2+\Delta)
\ge
 F_{N(\Delta)}(0,\Delta)+R_{N(\Delta),\Delta}(\beta).  \end{eqnarray}
 As follows from Proposition 2.7 of \cite{cf:GT_ALEA}, there exists
 $a_8\in(0,\infty)$ (depending only on the law $K(\cdot)$ of the
 renewal) such that \begin{eqnarray}
\label{eq:alea}
  F_N(0,\Delta)\ge F(0,\Delta)-a_8\frac{\log N}N
  \end{eqnarray}
  for every $N$.    Choosing $c=c(\epsilon)$ large
 enough, Eq. \eqref{eq:alea} implies that
 \begin{eqnarray}
 \label{eq:fsc}
   F_{N(\Delta)}(0,\Delta)\ge (1-\epsilon)F(0,\Delta).
 \end{eqnarray}
 As for $R_{N(\Delta),\Delta}(\beta)$, we have from \eqref{eq:integrating}
and \eqref{eq:315}
 \begin{equation}
 \begin{split}
  (1-e)^{-1}R_{N(\Delta),\Delta}(\beta)\,  \le &\,  F(0,\Delta)
\left(\frac1q\frac{F(0,q\Delta)}{F(0,\Delta)}-1\right)
+\epsilon F(0,\Delta)\\
& \phantom{moveright}
+
 \frac1{2N(\Delta)p}\log \bE^{\otimes 2} 
\left(e^{2p\beta^2\sum_{n=1}^{N(\Delta)}
 \delta^{(1)}_n\delta^{(2)}_n}
 \right),
 \end{split}
 \end{equation}
where we used Eqs. \eqref{eq:fsc} and \eqref{eq:superadd} to bound
$(1/q)F_{N(\Delta)}(0,q\Delta)-F_{N(\Delta)}(0,\Delta)$ from above.
Choosing again $q=q(\epsilon)$ we
obtain, for  $\Delta\le \Delta_0(\epsilon)$, 
 \begin{eqnarray}
\label{eq:aus1}
  (1-e)^{-1}R_{N(\Delta),\Delta}(\beta)\le 2\epsilon F(0,\Delta) +
\frac1{2N(\Delta)p(\epsilon)}\log \bE^{\otimes 2}
\left(e^{2p(\epsilon)\beta^2\sum_{n=1}^{N(\Delta)}
\delta^{(1)}_n\delta^{(2)}_n} \right).
 \end{eqnarray}
 Now observe that, if $1/2<\alpha<1$, there exists
 $a_{9}=a_{9}(\alpha)\in(0,\infty)$ such that for every integers $N$ and $k$
 \begin{eqnarray}
 \label{eq:intersec1}
   \bP^{\otimes 2}\left(\sum_{n=1}^N
\delta^{(1)}_n\delta^{(2)}_n\ge k\right)\le
 \left(1-a_{9}\frac{L(N)^2}{N^{2\alpha-1}}\right)^k.
 \end{eqnarray}
This geometric bound is proven in \cite[Lemma 3]{cf:Ken}, but
in Subsection~\ref{sec:alternate}  we give another
 simple proof.
Thanks to  \eqref{eq:intersec1} we have
\begin{eqnarray}
\label{eq:thanks}
  \bE^{\otimes 2} \left(e^{2p(\epsilon)\beta^2\sum_{n=1}^{N(\Delta)}
\delta^{(1)}_n\delta^{(2)}_n} \right)\,\le\,
\left(
{1-e^{2\beta^2p(\epsilon)} \left(1-a_{9}\frac{L(N(\Delta))^2}
{N(\Delta)^{2\alpha-1}}\right)}\right)^{-1},
\end{eqnarray}
whenever the right-hand side is positive, and this is of
course the case under the stronger requirement
\begin{eqnarray}
\label{eq:stronger}
e^{2\beta^2p(\epsilon)}
\left(1-a_{9}\frac{L(N(\Delta))^2}{ N(\Delta)^{2\alpha-1}}\right)\, \le\,
\left(1-\frac{a_{9}}2 \frac{L(N(\Delta))^2}{N(\Delta)^{2\alpha-1}}\right).
\end{eqnarray}
At this point, using the definition of $N(\Delta)$, it is not
difficult to see that there exists a positive constant
$a_{10}(\epsilon)$ such that
\eqref{eq:stronger} holds if
\begin{eqnarray}
\label{eq:horr}
\beta^2 p(\epsilon) \, &\le& \, a_{10}(\epsilon) \Delta^{(2\alpha -1)/\alpha}
\hat L(1/\Delta)\\\nonumber
&:=& a_{10}(\epsilon) \Delta^{(2\alpha -1)/\alpha}
\left[ \frac{
\tilde L(1/\Delta)}
{\left \vert \log F(0, \Delta) \right\vert}
\right] ^{2\alpha -1}\left(L\left(\frac{|\log F(0, \Delta)|}{ F(0, \Delta)}
\right)\right)^2.
\end{eqnarray}
The fact that $\hat L(\cdot)$ is slowly varying follows from
\cite[Proposition 1.5.7]{cf:bingh} and Eq. \eqref{eq:asint}. For
instance, if $L(\cdot)$ is asymptotically constant one has $\hat
L(x)\sim a_{11}|\log x|^{1-2\alpha}$. Condition \eqref{eq:horr} is
equivalent to the first inequality in \eqref{eq:condiz}, for suitably
chosen $a_1(\epsilon)$ and $\check L(\cdot)$ . As a consequence,
\begin{eqnarray}
\label{eq:asa}
   \frac1{2N(\Delta)p(\epsilon)}
\log \bE^{\otimes 2} \left(e^{2p(\epsilon)\beta^2\sum_{n=1}^{N(\Delta)}
 \delta^{(1)}_n\delta^{(2)}_n}
 \right)\le
\frac{F(0,\Delta)}{2c(\epsilon)p(\epsilon)|\log F(0,\Delta)|}
\log\left(\frac{2N(\Delta)^{2\alpha-1}}{a_{9} L(N(\Delta))^2}
\right).
\end{eqnarray}

Recalling Eq. \eqref{eq:asint} one sees that, if $c(\epsilon)$ is
chosen large enough,
\begin{eqnarray}
\label{eq:together}
   \frac1{2N(\Delta)p(\epsilon)}
\log \bE^{\otimes 2} \left(e^{2p(\epsilon)\beta^2\sum_{n=1}^{N(\Delta)}
 \delta^{(1)}_n\delta^{(2)}_n}
 \right)\le \epsilon F(0,\Delta).
\end{eqnarray}
Together with Eqs. \eqref{eq:identity2},
\eqref{eq:fsc} and \eqref{eq:aus1}, this concludes the proof of the
theorem.
\hfill $\Box$

\medskip

{\sl Proof of Theorem \ref{th:a32}}. The proof is almost identical to
that of Theorem \ref{th:a322} and up to Eq. \eqref{eq:aus1} no changes
are needed. The estimate \eqref{eq:intersec1} is then replaced by
 \begin{eqnarray}
\label{eq:intersec2}
\bP^{\otimes
 2}\left(\sum_{n=1}^N \delta^{(1)}_n\delta^{(2)}_n\ge k\right)\le
 \left(1-\frac{a_{12}}{\ell(N)}\right)^k.  \end{eqnarray} for every $N$,
 for some $a_{12}>0$ (see \cite[Lemma 3]{cf:Ken}, or the alternative argument
 given in Subsection~\ref{sec:alternate}).
In analogy with Eq. \eqref{eq:thanks} one obtains then
\begin{eqnarray}
\label{eq:thanks2}
  \bE^{\otimes 2} \left(e^{2p(\epsilon)\beta^2\sum_{n=1}^{N(\Delta)}
\delta^{(1)}_n\delta^{(2)}_n} \right)\le
\left(1-e^{2\beta^2p(\epsilon)} \left(1-\frac{a_{12}}{\ell(N(\Delta))}\right)
\right)^{-1}
\end{eqnarray}
whenever the right-hand side is positive.  Choosing $a_2(\epsilon)$
large enough one sees that if condition
\eqref{eq:condiz3} is fulfilled then
\begin{eqnarray}
e^{2\beta^2p(\epsilon)}\left(1-\frac{a_{12}}{\ell(N(\Delta))}\right)\le
\left(1-\frac{a_{12}}{2\ell(N(\Delta))}\right)
\end{eqnarray}
and, in analogy with \eqref{eq:asa},
\begin{eqnarray}
   \frac1{2N(\Delta)p(\epsilon)}
\log \bE^{\otimes 2} \left(e^{2(\epsilon)\beta^2\sum_{n=1}^{N(\Delta)}
 \delta^{(1)}_n\delta^{(2)}_n}
 \right)\le
\frac{F(0,\Delta)}{2c(\epsilon)p(\epsilon)|\log F(0,\Delta)|}
\log\left(\frac{2\ell(N(\Delta))}{a_{12}}\right).
\end{eqnarray}
From this estimate, for $c(\epsilon)$ sufficiently large one obtains
again \eqref{eq:together} and as a consequence the statement of
Theorem \ref{th:a32}.
\hfill $\Box$

\subsection{Proof of \eqref{eq:intersec1} and \eqref{eq:intersec2}}
\label{sec:alternate}
For what concerns \eqref{eq:intersec1},
start from the obvious bound
 \begin{eqnarray}
\label{eq:indid}
     \bP^{\otimes 2}\left(\sum_{n=1}^N
 \delta^{(1)}_n\delta^{(2)}_n\ge k\right)\le\left(
 1-\bP^{\otimes 2}(\inf\{n>0:n\in\tau^{(1)}\cap \tau^{(2)}\}>N)\right)^k.
 \end{eqnarray}
 Next note that, by Eq. \eqref{eq:doney},
 \begin{eqnarray}
\label{eq:asu}
   u_n:=\bP^{\otimes 2}(n\in \tau^{(1)}\cap\tau^{(2)})\stackrel{n\to\infty}
\sim \frac{C_\alpha^2}
 {L(n)^2 n^{2(1-\alpha)}}
 \end{eqnarray}
 and that $u_n$ satisfies the renewal equation
 \begin{eqnarray}
 \label{eq:reneq}
 u_n=\delta_{n,0}+ \sum_{k=0}^{n-1}u_k Q(n-k)
 \end{eqnarray}
 where $Q(k):=\bP^{\otimes 2}(\inf\{n>0:n\in\tau^{(1)}\cap \tau^{(2)}\}=k)$
 is the probability we need to estimate in \eqref{eq:indid}.
$Q(\cdot)$ is a probability
 on $\N$ since the renewal
$\tau^{(1)}\cap \tau^{(2)}$ is recurrent for $1/2<\alpha<1$, as
can be seen from the fact that, due to Eq.
\eqref{eq:doney}, the expectation in
\eqref{eq:ricorrenza} diverges in this case.
 After a Laplace transform, one finds for $s>0$
 \begin{eqnarray}
\label{eq:laplace}
 \hat Q(s):=\sum_{n\ge0}e^{-n s}Q(n)=1-\frac1{\hat u(s)}
 \end{eqnarray}
 and, by \cite[Theorem 1.7.1]{cf:bingh} and the asymptotic behavior
\eqref{eq:asu}, one finds
\begin{eqnarray}
\label{eq:asLap}
  \hat u(s)\stackrel{s\to 0^+}\sim \frac{C_\alpha^2
\Gamma(2\alpha)}{2\alpha-1}\frac1{s^{2\alpha-1}(L(1/s))^2}.
\end{eqnarray}
Note that $0<2\alpha-1<1$. By the classical Tauberian theorem
(in particular,
\cite[Corollary 8.1.7]{cf:bingh} is enough in this case), one obtains then
\begin{eqnarray}
  \sum_{n\ge N} Q_n\stackrel{N\to\infty}
\sim \frac{2\alpha-1}{C_\alpha^2\Gamma(2\alpha)
\Gamma\left(2(1-\alpha)\right)}
\frac{L(N)^2}{N^{2\alpha-1}}
\end{eqnarray}
which, together with \eqref{eq:indid}, completes the proof of
\eqref{eq:intersec1}.
\smallskip

We turn now to the proof of \eqref{eq:intersec2}.
From
 Eq. \eqref{eq:asu} with $\alpha=1/2$ and \cite[Theorem
 1.7.1]{cf:bingh}, one finds, in analogy with \eqref{eq:asLap},
\begin{eqnarray}
\label{eq:asLap2}
  \hat u(s)\stackrel{s\to 0^+}\sim C_{1/2}^2\ell (1/s).
\end{eqnarray}
Then, Eq. \eqref{eq:laplace} and \cite[Corollary 8.1.7]{cf:bingh}
imply
\begin{eqnarray}
\sum_{n\ge N}Q_n\stackrel{N\to\infty}\sim \frac 1{C_{1/2}^2\ell(N)}
\end{eqnarray}
and therefore Eq. \eqref{eq:intersec2}. \qed

\subsection{Proof of Theorem \ref{th:rs}}
Start again from \eqref{eq:identity} and define, 
for $q\in \R$,
\begin{eqnarray}
\phi_{N,\Delta}(t,\beta):=\frac1N\bbE \log 
\med{e^{\sum_{n=1}^N[\beta\sqrt t\go_n
-t\beta^2/2+\beta^2 q(t-1)]\delta_n}}_{N,\Delta}
\end{eqnarray}
so that
\begin{eqnarray}
\phi_{N,\Delta}(0,\beta)=F_N(0,\Delta-\beta^2q)-F_N(0,\Delta)
\end{eqnarray}
and $\phi_{N,\Delta}(1,\beta)=R_{N,\Delta}(\beta).$ In analogy with
Eq. \eqref{eq:ipp1} one has
\begin{eqnarray}
\nonumber
\frac{\dd}{\dd t}\phi_{N,\Delta}(t,\beta)&=&
-\frac{\beta^2}{2N}
\sum_{m=1}^N\bbE\left\{ \left(\frac{\med
{\delta_m\,
e^{\sum_{n=1}^N[\beta\sqrt t\go_n-t\beta^2/2+\beta^2 q(t-1)]
\delta_n}}_{N,\Delta}}
{\med
{e^{\sum_{n=1}^N[\beta\sqrt t\go_n-t\beta^2/2+\beta^2 q(t-1)]
\delta_n}}_{N,\Delta}}-q\right)^2
\right\}+\frac{\beta^2q^2}2\\
&\le&\frac{\beta^2q^2}2,
\end{eqnarray}
from which statement \eqref{eq:RS} follows after an integration on
$t$ (it is clear that taking the infimum over $q\in \R$ or over
$0\le q\le \Delta/\beta^2$ gives the same result.) The strict inequality in
\eqref{eq:RS} holds since the quantity to be minimized in \eqref{eq:RS} 
has negative derivative at $q=0$.

To prove \eqref{eq:svil} recall that $F(0,\Delta)$ satisfies for
$\Delta>0$ the identity
\cite[Appendix A]{cf:GT_smooth}
\begin{eqnarray}
\sum_{n\in\N}e^{-F(0,\Delta)n}K(n)=e^{-\Delta},
\end{eqnarray}
(so that, in particular, $F(0,\Delta)$ is real analytic for
$\Delta>0$). An application of \cite[Theorem 1.7.1]{cf:bingh}
gives therefore, for $\alpha<1/2$,
\begin{eqnarray}
\label{eq:L()}
\partial_\Delta F(0,\Delta)=\Delta^{(1-\alpha)/\alpha} L^{(1)}(1/\Delta),
\;\;\;\;
\partial^2_\Delta F(0,\Delta)=\Delta^{(1-2\alpha)/\alpha}L^{(2)}(1/\Delta),
\end{eqnarray} 
where the slowly varying functions $ L^{(i)}(\cdot)$ can be expressed
through $L(\cdot)$ (cf., for instance, \cite[Section 2.4]{cf:GB} for
the first equality). For $\alpha=0$, \eqref{eq:L()} is understood to
mean that the two derivatives vanish faster than any power of
$\Delta$.  This shows that $\partial^2_\Delta F(0,\Delta)$ is bounded
above by a constant for, say, $\Delta\le1$ if $\alpha<1/2$.  Then,
choosing $q=\partial_\Delta F(0,\Delta)$ in \eqref{eq:RS} (which is
the minimizer of $\beta^2q^2/2+F(0,\Delta-\beta^2q)$ at lowest order
in $\beta$) yields
\eqref{eq:svil}. It is important to note that, thanks to 
the first equality in \eqref{eq:L()} and the assumption 
$\alpha<1/2$,  this choice is compatible
with the constraint $q\le \Delta/\beta^2$, for $\Delta$ and
$\beta$ sufficiently small.
\qed

\begin{rem}\rm
\label{rem:why?}
The reason why we stopped at order $\beta^2$ in \eqref{eq:svil} is that
at next order the error term $O(\beta^6)$ involves
$\partial^3_\Delta F(0,\Delta)$, which diverges for $\Delta\searrow 0$ if
$\alpha>1/3$.  In analogy with \eqref{eq:svil}, one can however prove
that, if $\alpha<1/k$ with $2<k\in \N$, the expansion \eqref{eq:svil}
can be pushed to order $\beta^{2(k-1)}$ with a uniform control in
$\Delta$ of the error term $O(\beta^{2k})$.  We do not detail this
point,  the computations involved being straightforward.
\end{rem}

\section*{Acknowledgments}
I am extremely grateful to Giambattista Giacomin for many motivating 
discussions and for constructive comments on this manuscript.
This research has been conducted in the framework of the GIP-ANR
project JC05\_42461 ({\sl POLINTBIO}).

\end{document}